\PassOptionsToPackage{pdfpagelabels=false}{hyperref} 
\documentclass[fleqn,usenatbib,useAMS]{mnras}

\usepackage[T1]{fontenc}
\usepackage{ae,aecompl}
\usepackage{times} 
\usepackage{comment}
\usepackage{ulem}
\usepackage{graphicx}	
\usepackage{amsmath}	
\usepackage{amssymb}	
\usepackage{color}
\usepackage{upgreek} 
\usepackage[utf8]{inputenc}
\usepackage{slashed}
\usepackage{mathtools}

\makeatletter
\newcommand{\vast}{\bBigg@{4}}
\newcommand{\Vast}{\bBigg@{5}}

\makeatother

\newcommand{\e}{{\rm e}}
\newcommand{\dd}{{\rm d}}
\newcommand{\ii}{{\rm{i}}}
\newcommand{\fett}[1]{\boldsymbol{#1}}
\newcommand{\be}{\begin{equation}}
\newcommand{\ee}{\end{equation}}
\newcommand{\D}{\upartial_\tau^{\rm L}}
\newcommand{\R}{\mathfrak{R}_{\tau}}

\title[Shell-crossing in quasi-1D flow]{Shell-crossing in quasi-one-dimensional flow}

\author[C.\ Rampf and U.\ Frisch]{Cornelius Rampf$^{1,2}$\thanks{E-mail: \href{mailto:rampf@thphys.uni-heidelberg.de}{rampf@thphys.uni-heidelberg.de}}
  and Uriel Frisch$^{3}$ \\
$^{1}$Institute for Theoretical Physics (ITP), Philosophenweg 16, University of Heidelberg,  D-69120 Heidelberg, Germany \\
$^{2}$Department of Physics, Israel Institute of Technology -- Technion, Haifa 32000, Israel \\
$^{3}$Lab.\ Lagrange, UCA, OCA, CNRS, CS 34229, F-06304 Nice Cedex 4, France 
}

\date{\today}

\pubyear{2017}

\begin{document}

\label{firstpage}
\pagerange{\pageref{firstpage}--\pageref{lastpage}}
\maketitle

\begin{abstract}
  Blow-up of solutions for the cosmological fluid equations, often dubbed shell-crossing or orbit crossing, denotes the breakdown of the single-stream regime of the cold-dark-matter fluid. At this instant, the velocity becomes multi-valued and the density singular. Shell-crossing is well understood in one dimension (1D), but not in higher dimensions. This paper is about quasi-one-dimensional (Q1D) flow that depends on all three coordinates but differs only slightly from a strictly 1D flow, thereby allowing a perturbative treatment of shell-crossing using the Euler--Poisson equations written in Lagrangian coordinates. The signature of shell-crossing is then just the vanishing of the Jacobian of the Lagrangian map, a regular perturbation problem. In essence the problem of the first shell-crossing, which is highly singular in Eulerian coordinates, has been desingularized by switching to Lagrangian coordinates, and can then be handled by perturbation theory. Here, all-order recursion relations are obtained for the time-Taylor coefficients of the displacement field, and it is shown that the Taylor series has an infinite radius of convergence. This allows the determination of the time and location of the first shell-crossing, which is generically shown to be taking place earlier than for the un-perturbed 1D flow.

The time variable used for these statements is not the cosmic time $t$ but the linear growth time $\tau \sim t^{2/3}$. For simplicity, calculations are restricted to an Einstein--de Sitter universe in the Newtonian approximation, and tailored initial data are used. However it is straightforward to relax these limitations, if needed.
\end{abstract}


\section{Introduction}
\label{sec:intro}

It is widely known that exact analytic solutions to the cosmological
fluid equations exist for initial data that only depend on one space
variable.  These play an important role in cosmology, not only because
they are simple but because the breakdown of smooth three-dimensional (3D)
solution through the
development of infinite-density caustics begins generically as an almost
1D phenomenon with the formation of pancakes \citep{MS:1989, MS:1993}.
When the problem is exactly 1D,
the fluid equations become linear when expressed in Lagrangian
coordinates.  As a consequence, the linear Lagrangian solution solves
the fully non-linear problem in 1D \citep{Novikov:2010ta,ZentsovaChernin1980}.  
Expressed in comoving coordinates and enabling the linear growth time $\tau$
(see below for details), the resulting Lagrangian map is in one
spatial dimension exactly
\be \label{eq:ZAintro}
 {x}({q};\tau) = {q} + \tau\, {v}^{\rm (init)}({q}) \,,
\ee
where ${q}$ and ${v}^{\rm (init)}$ are respectively the initial
position and velocity of the fluid particle.  The last term on the
right-hand side in equation~\eqref{eq:ZAintro} is the one-dimensional
Lagrangian displacement field. It is linear in the time variable, and,
evidently, the 1D displacement could be viewed as the first and only
non-zero term of an infinite time-Taylor series around $\tau=0$.
Obviously, the 1D Lagrangian map \eqref{eq:ZAintro} is analytic in
$\tau$, has no singularities and thus an infinite radius of
convergence. Singularities appear however when reverting back to
Eulerian coordinates, since the Lagrangian map is not invertible
anymore when its Jacobian vanishes for the first time.
At this instant, commonly referred to as shell-crossing, 
the fluid enters the multi-stream regime, which implies
that the single-stream fluid description breaks down 
(in both Lagrangian and Eulerian coordinates).

Generally, departing from 1D leads to a non-zero population of 
higher-order time-Taylor coefficients of the displacement that should be taken into account.
For generic 3D initial data, low-order solutions of the displacement are well known, 
see for example the first-order solution, which is called the 
Zel'dovich approximation
\citep[ZA;][]{Zeldovich:1969sb}; 
for a generalization of this approximation, see \citet{Buchert:1992ya}.
Explicit solutions 
to the second \citep{Buchert:1993xz,Bouchet:1992uh}, 
to the third \citep{Buchert:1993ud,Bouchet:1994xp}, 
and to the fourth order \citep{Rampf:2012xa} are known as well.
Furthermore, truncated approximations for the 3D displacement up to the third
order have been applied to numerically extrapolate 
for the particle trajectories, see e.g.,   
\citet{Buchert:1991ez,Buchert:1993df,Melott:1994ah,Buchert:1995km,Tassev:2012cq}.

For generic 3D initial data, the radius of convergence for the time-Taylor series of the 
displacement field in Lagrangian coordinates is, most likely, not infinite anymore but 
determined by complex-time singularities, not related to shell-crossing.
Analytical bounds on the radius of convergence 
can be obtained by investigating the large-order behaviour of the infinite time-Taylor
series. This, of course, requires explicit all-order recursion relations, as obtained by 
 \citet{Zheligovsky:2013eca}, who used their 
recursion relations for the displacement to obtain a 
lower bound on the radius of convergence. Such a lower bound amounts to 
finding a time $T$, as large as possible, such that the time-Taylor 
series around $\tau=0$ is guaranteed to converge for
$0 \leq \tau \leq T$.
Furthermore, \citet{Rampf:2015mza} have shown that shell-crossing is ruled 
out in that time-domain.
Obtaining the time of first shell-crossing with generic 3D initial data
can probably be investigated only by numerical means, for example by
employing the multi-time stepping algorithm called the Cauchy--Lagrangian method, 
an algorithm that has so far been implemented only for incompressible flow \citep{CL2016}.

In the present paper, we show that much more can be handled analytically
when restricting the initial data to being close to 1D. 
We prove that the relevant Lagrangian map has infinitely many non-vanishing terms 
in its time-Taylor series, but that the series is entire in time, that is, it has 
an infinite radius of convergence. Thus particle trajectories can be evaluated
in a single time-step from initial time all the way up to the first shell-crossing 
(but not beyond, for reasons that will be discussed later).
To unravel the above, we make use of novel recursion relations 
which are, most importantly and crucially, 
tailored to initial data that are perturbatively close 
to one-dimensional. By contrast,
all-order recursive solutions for generic 3D initial data  
\citep[see also][]{Goroff:1986ep,EhlersBuchert97,Rampf:2015mza,Matsubara:2015ipa} 
are not suitable for the considered problem. Indeed,  the usage of Q1D initial data
introduces another expansion parameter, namely a parameter which parametrizes
the perturbative departure from 1D. 
Thus, the power counting in the perturbative expansion in Q1D
is formally different from the generic 3D case.
For this reason, our results are not contained in the commonly used Lagrangian perturbation solutions.

This paper is organized as follows. In section~\ref{sec:EPequations} we review the 3D 
Euler--Poisson equations, 
first in the \mbox{Eulerian- and} then in the Lagrangian-coordinates approach.
The latter approach serves as our starting point for the present paper.
 In section~\ref{sec:1+eps}, we  show how to embed the Q1D problem 
into three-dimensional space, and particularly discuss the used initial conditions
and perturbation \textit{Ansatz}.
The resulting equations can be easily solved to a given order in the book-keeping
perturbation variable $\epsilon$. The zeroth-order solution in~$\epsilon$, which we
call the solution of the {\it unperturbed problem}, is the one for
which the initial 
data depends only on one space variable (i.e., the 1D case). The first-order equations 
to \mbox{order $\epsilon$,}
which resemble the perturbed equations with respect to the unperturbed problem, 
are given in section~\ref{sol:e1}.
We solve this {\it perturbed problem} by using a time-Taylor series in 
section~\ref{sec:Taylor}. The proof of the absence of singularities in the perturbed 
Lagrangian equations 
is given in section~\ref{sec:nosing}.  
In section~\ref{sec:tstar} we show how to obtain the  time and location 
of the first shell-crossing. In section~\ref{3waves} we give a concrete 
example involving a three-sine wave Q1D initial condition.
Concluding remarks are presented in section~\ref{sec:conclusions}.

\newpage

\section{Euler--Poisson equations in 3D}
\label{sec:EPequations}

\subsection{Basic equations in Eulerian coordinates}\label{sec:Euler}

The Euler--Poisson equations 
are usually formulated in comoving coordinates $\fett{x}= \fett{r}/a$, where
$\fett{r}$ is the proper space coordinate and 
$a$ the cosmic scale factor. The latter parametrizes
the global background/Hubble expansion, and its
evolution is given by the usual Friedmann equations.
In the present work, we restrict our analysis, for simplicity, 
to an Einstein--de Sitter (EdS) cosmology,
where the universe is filled only with a cold dark matter (CDM) fluid;
the cosmological constant, usually denoted by $\Lambda$, is set to
zero. This and many other approximations are however easily rectified if needed, see
e.g.\ \citet{Rampf:2015mza} for an analysis within the $\Lambda$CDM model and beyond.

We denote by $\fett{v}$ the peculiar velocity with respect to the Hubble flow, 
by $\delta = (\rho - \bar \rho)/\bar \rho$ the density contrast with background density $\bar \rho \sim a^{-3}$,
by $\varphi_{\rm g}$ the cosmological potential, and by $\tau$ the 
{\it linear growth time} (often denoted with $D$ or $D(t)$). 
For an EdS universe, $\tau$ is related to the cosmic time $t$ via $\tau \sim t^{2/3}$.
The Euler--Poisson equations for an EdS universe are 
\citep[\citealt{Brenier:2003xs}; for more general cosmologies see][]{1992STIN...9519341S,1994PhyD...77..342S}
\begin{subequations} \label{fluidequations}
\begin{align}
 &\upartial_\tau \fett{v} + (\fett{v} \cdot \nabla) \fett{v} = -\frac{3}{2\tau} \left( \fett{v} + \nabla \varphi_{\rm g}\right) \,,  \label{eq:Euler} \\
 &\upartial_\tau \delta + \nabla \cdot \left[ (1+\delta) \fett{v} \right] = 0 \,, \\
 &\nabla^2 \varphi_{\rm g} = \frac \delta \tau \,, \label{eq:Poisson}
\end{align}
\end{subequations}
(here $\nabla^2$ is the Eulerian Laplacian). Enabling $\tau$ as the time variable 
is convenient when studying the well-posedness of the fluid equations at  short times, 
as explained hereafter. It is actually essential when investigating
the time-analyticity of the Lagrangian map \citep{Zheligovsky:2013eca,Rampf:2015mza}.

Formally linearizing around the steady state $\fett{v} = \fett{0}$ and $\delta =0$, 
the above equations can be written in terms of a single differential equation for 
the density contrast \citep{Peebles1980},
\be
 \upartial_{\tau \tau}^2 \delta = - \frac{3}{2\tau} \left( \upartial_\tau \delta - \frac \delta \tau \right) .
\ee
This equation has two power-law solutions. One is called the decaying mode which 
behaves as $\tau^{-3/2}$ and thus blows up when $\tau \to 0$, 
thereby invalidating linearization. 
The other one is linear in  $\tau$ and thus trivially analytic; 
furthermore it stays analytic in the presence of a 
non-vanishing cosmological constant \citep{Rampf:2015mza}.  
Therefore,
$\tau$ is  the physically appropriate variable  
for describing the growth of density fluctuations at short times.

Before investigating the fully non-linear theory in Lagrangian coordinates, 
let us briefly discuss an important feature of the Euler--Poisson equations. 
Observe the presence of the linear growth time in the denominators of the 
right-hand side of equations~(\ref{eq:Euler}) and~(\ref{eq:Poisson}).
This indicates that the solution at $\tau=0$ is singular unless the 
following {\it slaving conditions} are satisfied [denoted by the superscript (init)]:
\be \label{slaving}
  \delta^{(\rm init)} = 0 \,, \qquad \fett{v}^{(\rm init)}  = - \nabla \varphi_{\rm g}^{(\rm init)}\,. 
\ee
As argued by \citet{Rampf:2015mza}, the second condition immediately implies that the velocity 
is potential at initial time, a feature that, because of~(\ref{eq:Euler}), 
persists also at later times in Eulerian coordinates,
\be \label{novorticity}
  \nabla \times \fett{v} = \fett{0}  \,.
\ee
(In Lagrangian coordinates, the potential character of the velocity is generally lost.)
Finally, we note that in our mathematical description, we allow the time variable $\tau$ 
to become arbitrarily small. From the point of view of boundary layer analysis, this amounts 
to subsuming the whole primordial physics into the slaving conditions; 
see \citet{Zheligovsky:2013eca,Rampf:2015mza} for detailed discussions.
Nevertheless, initialization times with $\tau^{\rm (init)}>0$ can be incorporated 
in our formalism which however requires some additional calculational steps; 
we shall come back to this issue in section~\ref{sec:conclusions}.

\subsection{Basic equations in Lagrangian coordinates}

We denote by $\fett{q}$ the Lagrangian coordinates with components
$q_i$ ($i$=1,2,3); a partial derivative with respect to $q_i$ acting
on a given function $f$ is denoted by $f_{,i}$ and, occasionally, by $\upartial_i ^{\rm L}
f$. Summation over repeated indices is implied, and, for simplicity,
since we work in the Newtonian limit, we do not distinguish between
contra- and covariant coordinate indices.
Let $\fett{q} \mapsto {\fett{x}}(\fett{q};\tau)$ 
be the Lagrangian map from the initial ($\tau\!=\!0$)
position $\fett{q}$ to the Eulerian position ${\fett{x}}$ at \mbox{time $\tau$.}
The map satisfies $\fett{v}(\fett{x}(\fett{q};\tau);\tau) = \D \fett{x}(\fett{q};\tau)$,  
where $\D$ is the Lagrangian time derivative -- the latter also denoted with an overdot in the following.
At initial time, $\tau=0$, the velocity is
\be
 \fett{v}^{\rm (init)}(\fett{q}) = \fett{v}(\fett{x}(\fett{q};0);0) \,,
\ee
which agrees with the initial Eulerian velocity.
Mass conservation is, until the first shell-crossing, given by 
\be  \label{lagmass}
 \delta =1/J -1 \,,
\ee 
where $J= \det( x_{i,j} )$, the determinant of the Jacobian matrix with entries $x_{i,j}$, 
is called the Jacobian (as long as it is non-negative). With these definitions, 
the Euler--Poisson equations can be written in Lagrangian coordinates in the compact form 
\begin{subequations} \label{eqs:lag}
\begin{align}
 &\varepsilon_{ikl} \varepsilon_{jmn} \, x_{k,m}\, x_{l,n} \R x_{i,j} = 3 \left( J-1 \right) \,, \label{eq:scalarLag} \\
 &\varepsilon_{ijk} \, \dot x_{l,j} x_{l,k} = 0 \,, \label{eq:Cauchy} 
\end{align}
\end{subequations}
where we have defined the operator $\R \equiv \tau^2 \left(\upartial_{\tau}^{\rm L}\right)^2 + (3\tau/2) \D$, 
and $\varepsilon_{ijk}$ is the fundamental antisymmetric tensor. 
Equation~(\ref{eq:scalarLag}) is a scalar equation that results from combining equations~(\ref{eq:Euler}) 
and~(\ref{eq:Poisson}), and by taking mass conservation~\eqref{lagmass} into 
account \cite[for a derivation see, e.g.,][and in there set $\Lambda=0$]{Rampf:2015mza}.
Equations~\eqref{eq:Cauchy} are the {\it Cauchy invariants;} these are Lagrangian (kinematical) 
constraints on the Lagrangian map that must be satisfied in order to maintain the curlfree motion in Eulerian space.
The Cauchy invariants can thus be understood as the 
corresponding Lagrangian counterpart of equation~\eqref{novorticity}.
See e.g.\ \citet{Rampf:2016wom} for a detailed derivation of the Cauchy invariants.

Equations~\eqref{eqs:lag} constitute the well-known closed system of Lagrangian equations for CDM. 
As can be easily checked, power-series solutions for these equations are non-singular at $\tau=0$,
 provided one makes use of the slaving conditions~\eqref{slaving}. 
Actually, such expansions in powers of the linear growth time $\tau$ are very common
in the Lagrangian perturbation theory 
[see e.g.\ \citet{Buchert:1993ud,Matsubara:2007wj,Rampf:2012xa,Zheligovsky:2013eca}].

\section{The quasi one-dimensional problem embedded in 3D}\label{sec:1+eps}

The aim of this paper is to analyse three-dimensional shell-crossing 
with initial conditions (ICs) that are close to one dimension, i.e., the 
ICs depend, to the zeroth order in a  perturbation parameter $\epsilon$, 
only on one space variable, and, 
to first order in $\epsilon$, in general on all space variables. 
Appropriate ICs and our perturbation \textit{Ansatz} are 
introduced in the following two sections.
Equations to \mbox{order $\epsilon^0$ and $\epsilon^1$} are then given 
in sections~\ref{sol:e0} and~\ref{sol:e1}, respectively.

\subsection{Initial conditions}\label{sec:IC}

Quasi one-dimensional initial conditions can be formulated in terms of a
superposition of two contributions for the initial gravitational potential. 
The first is an arbitrary function in the space variable $q_1$
and the second one  a small perturbation, proportional to $\epsilon$, 
which depends generally on all space variables.
Although the former, which characterizes the initial conditions 
for the purely one-dimensional problem, could be taken quite arbitrary (within
a class of function guaranteeing well-posedness for at least a finite time), 
it is advantageous to choose this function wisely: We know
that in the one-dimensional case, the occurrence of the first shell-crossing 
will appear downstream  at the spatial position $q_1$ where the initial velocity gradient
achieves its most negative value. By a suitable spatial translation, we can
take this location to be $q_1=0$. Then, by a suitable Galilean transformation,
we can take the velocity at this location to be zero. An instance is to take for this
one-dimensional gravitational potential the function $- \cos q_1$. 
Considerations of normal-form reduction indicate that we can actually make this choice
of initial conditions without loss of generality (within the class of $2\uppi$-periodic functions). 

As to  the perturbation, it  must be taken fairly general.  We thus use the initial data
\be \label{initialdata}
  \varphi_{\rm g}^{(\rm init)}(q_1,q_2,q_3) =  -\cos q_1 + \epsilon \,\phi^{\rm (init)}(q_1,q_2,q_3) \,,
\ee
where $\epsilon>1$ is a small perturbation parameter and $\phi^{\rm (init)}$ an 
arbitrary $2\uppi$-periodic function of $(q_1,q_2,q_3)$. Without loss of generality, 
we can take  the transverse average to be zero, namely
\be 
 \langle \phi^{\rm (init)} \rangle \equiv 
\int_0^{2\uppi} \frac{\dd q_2}{2\uppi} \int_0^{2\uppi} \frac{\dd q_3}{2\uppi} \, \phi^{\rm (init)}(q_1,q_2,q_3) =0\,. \label{zeroaverage}
\ee
Indeed, if this average is a non-trivial function of $q_1$, we can incorporate
it into the unperturbed flow.

Taking into account the equality of the initial gravitational and velocity
potentials (imposed by slaving), the initial velocity is, in index notation,
\be
 v_i^{(\rm init)}(\fett{q}) = - \delta_{i1} \sin q_1 -
 \epsilon \,\upartial_i ^{\rm L} \phi^{\rm (init)}(\fett{q}) \,, \label{initialvelocity}
 \ee 
where $\delta_{ij}$ is the Kronecker delta.

\subsection{The Lagrangian perturbation \textit{Ansatz}}\label{sec:epsAnsatz}

We are going to use a perturbation method in which  the solutions to the
Lagrangian equations~\eqref{eqs:lag} are expanded in powers of
the small parameter $\epsilon$. Namely, we look for a solution in which
the Lagrangian map is given by the perturbation \textit{Ansatz}
\be
  \fett{x}(\fett{q};\tau) 
   = \fett{q} + \fett{\xi}^{(0)}(\fett{q};\tau)+ 
\epsilon \,\fett{\xi}^{(1)}(\fett{q};\tau) +\epsilon^2 \fett{\xi}^{(2)}(\fett{q};\tau) +\dots\,,
\label{fettexpand}
\ee
where $\fett{\xi}^{(n)}(\fett{q})$ is the coefficient of $\epsilon^n$ 
in the expansion of the displacement $\fett{x}-\fett{q}$. For $\epsilon =0$,
we are back to  exactly one dimension;  the displacement  depends
only on $q_1$ and is in the direction of the first coordinate axis. 
We can thus write
\be
\xi_i^{(0)}(\fett{q})= \delta_{i1}F(q_1;\tau)\,.
\label{xi0f}
\ee
In this paper, we do not expand beyond first order in $\epsilon$ and, for
brevity, we write $\fett{\xi}^{(1)}(\fett{q};\tau) = \fett{\xi}(\fett{q};\tau)$.
Henceforth all calculations are extended only to first order in $\epsilon$.
For example, from~\eqref{fettexpand} and~\eqref{xi0f}, we have thus
\be
  x_i(\fett{q};\tau) 
   = q_i + \delta_{i1} F(q_1; \tau) + \epsilon\, \xi_i(\fett{q}; \tau)\,. \label{ansatzTraj}
\ee

From \eqref{ansatzTraj} it follows that the Jacobian matrix is given by
\be
x_{i,j} = \delta_{ij} + \delta_{i1}  \delta_{j1} F_{,1} +\epsilon\,\xi_{i,j}\,,
\label{jacmatrix}
\ee
so that, ignoring $O(\epsilon^2)$ terms, its determinant, the Jacobian, is given by
\be \label{jacobepsilon}
  J = 1 + F_{,1} + \epsilon \left( \xi_{1,1} + \xi_{2,2} + \xi_{3,3}\right) + \epsilon F_{,1} \left( \xi_{2,2} + \xi_{3,3} \right) \,.
\ee

 The vanishing of the Jacobian is evidence of shell-crossing. 
In the following we solve for the displacement and evaluate the Jacobian at a given order 
in the $\epsilon$-perturbation. To zeroth-order the problem is effectively one-dimensional, 
with the {\it unperturbed Jacobian}
$J^{(0)} = 1 + F_{,1}$. Since the unperturbed problem is embedded in the three-dimensional 
space \citep[see][]{Buchert:1987xy}, the first shell-crossing will occur at a given 
value $q_{1\star}^{(0)} =0$ (modulo $2\uppi$)
and in the two-dimensional plane that is spanned by the $2$-tuple $(q_2, q_3)$.
Switching on the $\epsilon$-perturbation, the two-dimensional plane that marked the instance 
of shell-crossing in the unperturbed problem will collapse to a given point $(q_{1\star}, q_{2\star}, q_{3\star})$ (modulo $2\uppi$).
Also the time of the first shell-crossing will generally change when switching from the unperturbed 
to the perturbed problem. Analytical and calculational details to the respective values, 
$\tau_\star^{(0)}$ and $\tau_\star$, are discussed in the following sections.

\subsection{Lagrangian equations to zeroth order in \texorpdfstring{$\fett{\epsilon}$}{epsilon}}\label{sol:e0}

To lowest (zeroth) order in $\epsilon$, the problem is exactly
one-dimensional and has the initial condition $\varphi_{\rm g}^{(\rm init)}(q_1,q_2,q_3) =  -\cos q_1$.
In this instance, as is well known, the Zel'dovich approximation is exact until shell-crossing \citep{Novikov:2010ta,Zeldovich:1969sb}. In Lagrangian coordinates, this
is particularly obvious. Indeed, substituting
\textit{Ansatz}~(\ref{ansatzTraj}) into~(\ref{eq:Cauchy}) gives a trivial identity, whereas for equation~(\ref{eq:scalarLag}), we obtain
\begin{align}
 2 \R F_{,1} = 3 F_{,1} \,. 
\end{align}
This linear equation has, on the one hand, the decaying solution $\sim \tau^{-2/3}$, which
is unbounded as $\tau \to 0$ and not acceptable and, on the other hand the growing solution
\be \label{eq:ZA}
  F_{,1}(q_1;\tau) =  \tau\,\chi_{,1}(q_1) \,,
\ee
where $\chi(q_1)$ can be taken arbitrary, but by~(\ref{initialvelocity}) is actually equal to the
initial velocity 
\be
  \chi(q_1) = - \sin q_1 \,.
\ee
To zeroth order in $\epsilon$, the particle trajectory is thus
\be
  x_i = q_i -  \delta_{i1} \tau \sin q_1  \,,  \label{zerothorder}
\ee
and the unperturbed Jacobian is
\be
 J^{(0)} = 1 - \tau \cos q_1 \,.
\ee
The instance of first shell-crossing is determined by the lowest time-value for which
$ J^{(0)}( q_{1\star}^{(0)};\tau_\star^{(0)}) = 0$.
It is easily checked that this happens
at $\tau_\star^{(0)} = 1$ and within the two-dimensional $(q_2,q_3)$-plane 
at $q_{1\star}^{(0)} =  0$ (modulo $2\uppi$).

Note that the particle trajectory~\eqref{zerothorder} is a linear function in time~$\tau$. Hence its time-Taylor series 
has just two terms and, trivially, 
has an infinite radius of convergence; i.e.\ it is an entire function. A singularity for the one-dimensional
case appears only when reverting
to Eulerian coordinates. Indeed, when the Jacobian vanishes,  the Lagrangian map ceases to be smoothly and uniquely invertible.
In a purely Eulerian formulation, by contrast, infinities appear explicitly when the fluid enters the multi-stream regime. 
Actually, the Eulerian time-Taylor series for both the density and the
velocity have a singularity at shell-crossing. As we shall now see, this
picture remains basically unchanged, when we switch on the perturbation -- which
makes the problem fully three-dimensional.

\subsection{Lagrangian equations to first order in \texorpdfstring{$\fett{\epsilon}$}{epsilon}}\label{sol:e1}

Collecting all terms $O(\epsilon)$, we obtain from equation~(\ref{eq:scalarLag})
\begin{subequations} \label{eqs:epsilon}
\begin{align}
     \R \xi_{1,1} + \left[  1 - \tau \cos q_1  \right] \R \left(  \xi_{2,2}  + \xi_{3,3} \right) 
  = \frac 3 2 \xi_{l,l}  \,,\!\! \label{scalarEpsilon}
\end{align}
and from the three components of~(\ref{eq:Cauchy}), i.e.\ $i=1,2,3$,  respectively
\begin{align}
 & \dot \xi_{2,3} -  \dot \xi_{3,2}  =0 \,, \label{eq:epsCauchy1} \\
 & \!\left[ 1- \tau \cos q_1 \right] \dot \xi_{1,3} + \xi_{1,3} \cos q_1 =  \dot \xi_{3,1}  \,,  \label{eq:epsCauchy2} \\
& \!\left[ 1- \tau \cos q_1 \right] \dot \xi_{1,2} + \xi_{1,2} \cos q_1  = \dot \xi_{2,1} \,.  \label{eq:epsCauchy3}
\end{align}
\end{subequations}
Observe that equations~\eqref{eqs:epsilon} are, by construction, the Lagrangian
Euler--Poisson equations~\eqref{eqs:lag}, linearized around the exact 1D solution 
\eqref{zerothorder}. Since the latter depends explicitly on $q_1$, so do the 
linearized equations. But there is no explicit dependence on $q_2$ and $q_3$.
As a consequence, if we are able to solve equations~\eqref{eqs:epsilon}
 for the case where the
initial potential perturbation is $\varphi(q_1,q_2,q_3)= \tilde\varphi(q_1) \exp  \{ \ii (k_2q_2+k_3 q_3)\}$, 
i.e.\ with a single (transverse) Fourier
harmonic, then we can handle the general case by linear
superposition. In the single-harmonic case, derivatives with respect
to $q_2$ and $q_3$  can be replaced with $\ii k_2$ and $\ii k_3$, respectively.
Using this and defining
\be
   \chi \equiv \xi_1 \,, \qquad   \zeta \equiv \ii k_2 \xi_2 + \ii k_3 \xi_3 \,, \qquad k_\perp^2 \equiv k_2^2 + k_3^2 \,,
\ee
we are left with only two unknowns, and equations~\eqref{eqs:epsilon} reduce to just two equations, which are 
\begin{subequations} \label{evolutioneqs}
\begin{align}
  &\R \chi_{,1} + \left[  1 -  \tau \cos q_1   \right] \R \zeta 
   = \frac 3 2 \left(  \chi_{,1} + \zeta        \right) \,, \label{evo1} \\
 &k_\perp^2 \left\{ \left[ 1- \tau \cos q_1 \right] \dot \chi + \chi \cos q_1 \right\} =  - \dot \zeta_{,1} \,.  \label{evo2}
\end{align}
\end{subequations}
This is the basic  set of linearized Lagrangian equations, which we solve in the following section.

\section{Taylor expansion and recursion relations}\label{sec:Taylor}

We observe that the linearized equations \eqref{evolutioneqs}  constitute a system of two
linear partial differential equations in the variables $\tau$ (second order)  and $q_1$ (first order), 
in which the transverse coordinates appear only parametrically through the wavenumber $k_\perp$. 
Our method of solution will use time-Taylor expansions
to arbitrary high order, based on novel recursion relations for the Taylor
coefficients. 
 
For this, we seek a solution to~\eqref{evolutioneqs} 
in the form of a Taylor series in the $\tau$-time for the displacement components,
\be
  \chi(\fett{q};\tau) = \sum_{n=1}^\infty \chi^{(n)}(\fett{q}) \,\tau^n \,, \qquad 
  \zeta(\fett{q};\tau) = \sum_{n=1}^\infty \zeta^{(n)}(\fett{q})\, \tau^n \,.
\label{timetaylor}
\ee
Substituting this \textit{Ansatz} into equations~\eqref{evolutioneqs} and
collecting all the terms containing a given power in $\tau^n$, 
yields the following relations between the time-Taylor coefficients, 
\begin{subequations} \label{rec1}
\begin{align}
  &\!\!\left[  n^2 + \frac n 2 - \frac 3 2 \right] \left( \chi_{,1}^{(n)} + \zeta^{(n)} \right)  = \left[ n^2 - \frac{3n}{2} + \frac 1 2 \right] \zeta^{(n-1)} \cos q_1 , \\
 &n \,\chi^{(n)} + k_\perp^{-2} n\, \zeta_{,1}^{(n)} =  (n-2) \,\chi^{(n-1)} \cos q_1  \,.
\end{align}
\end{subequations}
Here and in the following, by construction, coefficients vanish if their index is zero or negative.
Equations~\eqref{rec1} can be simplified by Fourier transforming also in the $q_1$ variable. For this we define
\be
  \chi^{(n)} = \hat \chi_{k_1}^{(n)} {\rm e}^{\ii k_1 q_1} \,, \qquad  \zeta^{(n)} = \hat \zeta_{k_1}^{(n)} {\rm e}^{\ii k_1 q_1} \,,
\ee
and then, making use of Euler's formula $\cos q_1 = (\exp\{ \ii q_1\} + \exp\{ -\ii q_1 \})/2$, we obtain for equations~(\ref{rec1})
\begin{subequations} \label{rec2}
\begin{align}
  &\!\!\left[  n^2 + \frac n 2 - \frac 3 2 \right] \left( \ii k_1 \hat \chi_{k_1}^{(n)} + \hat \zeta_{k_1}^{(n)} \right)  = \left[  n^2 - \frac{3n}{2} + \frac 1 2 \right] \nonumber \\
    &\quad \hspace{3.8cm} \times \frac 1 2 \left( \hat \zeta_{k_1+1}^{(n-1)} + \hat \zeta_{k_1-1}^{(n-1)} \right) \,, \label{rec2.1} \\
 &n \,\hat \chi_{k_1}^{(n)} + k_\perp^{-2} n\, \ii k_1 \hat \zeta_{k_1}^{(n)} =  \frac{n-2}{2} \left( \hat \chi_{k_1+1}^{(n-1)} + \hat \chi_{k_1-1}^{(n-1)}  \right) \,. \label{rec2.2}
\end{align}
\end{subequations}
For $n=1$, the first of these equations amounts to an identity, 
and the last equation gives
\begin{align}
  \hat \chi^{(1)}_{k_1} &= - \ii k_1 \, k_\perp^{-2} \hat \zeta_{k_1}^{(1)} \,, 
\end{align}
and, together with the definition of the Lagrangian  map and~(\ref{initialvelocity}), thus
\begin{align}
  \hat \chi_{k_1}^{(1)} &= - \ii k_1 \hat \phi^{\rm (init)} \,,  \hspace{0.94cm}\hat \zeta_{k_1}^{(1)} = k_\perp^2 \hat \phi^{\rm (init)} \,. 
\end{align}
For $n>1$, we obtain from equations~(\ref{rec2}) the following explicit
recursion relations:
\begin{subequations}
\label{explicitrec}
\begin{align}\label{reczeta}
  &\hat \zeta_{k_1}^{(n)} = \left( 1 + \frac{k_1^2}{k_\perp^2} \right)^{\!-1}\!  \Bigg[  \frac{n-1/2}{2n+3} \left( \hat \zeta_{k_1+1}^{(n-1)} + \hat \zeta_{k_1-1}^{(n-1)} \right) \nonumber \\ 
   &\quad \hspace{2.6cm} - \ii k_1 \frac{n-2}{2n} \left( \hat \chi_{k_1+1}^{(n-1)} + \hat \chi_{k_1-1}^{(n-1)} \right)
  \Bigg] , \\ \label{recchi}
  &\hat \chi^{(n)}_{k_1} =   \left( 1 + \frac{k_1^2}{k_\perp^2} \right)^{\!-1}\! \Bigg[ - \ii k_1\, k_\perp^{-2} \frac{n-1/2}{2n+3} \left( \hat \zeta_{k_1+1}^{(n-1)} + \hat \zeta_{k_1-1}^{(n-1)} \right) \nonumber \\
   &\quad \hspace{2.6cm}  + \frac{n-2}{2n} \left( \hat \chi_{k_1+1}^{(n-1)}+\hat \chi_{k_1-1}^{(n-1)}\right)
   \Bigg] \,.
\end{align}
\end{subequations}
We then construct the  $n$th-order time-Taylor coefficient of the displacement field using a Helmholtz--Hodge decomposition. The latter 
reads in Fourier space ($n \geq 1$):
\be
  \hat{\fett{\xi}}_{k_1}^{(n)} = - \!\left( k_1^2 + k_2^2 + k_3^2 \right)^{-2} \!
    \left( \ii \fett{k} \left[  \ii \fett{k} \cdot \hat{\fett{\xi}}_{k_1}^{(n)}  \right]\! - \ii  \fett{k} \times \hat{\fett{T}}_{k_1}^{(n)}  \right) ,
\ee
with
\begin{align}
  & \ii \fett{k} \cdot \hat{ \fett{\xi}}_{k_1}^{(n)} = \ii k_1 \hat \chi_{k_1}^{(n)} + \hat \zeta_{k_1}^{(n)}  \,,  \\
 &\hat{\fett{T}}_{k_1}^{(n)} = \ii \left(0 , k_3, - k_2 \right)^{\rm T}
   \left(  \hat \chi_{k_1}^{(n)} + \ii k_1 k_\perp^{-2} \hat \zeta_{k_1}^{(n)} \right) \,.
\end{align}
The respective right-hand sides of the two last equations are 
combinations of time-Taylor coefficients to order $n$. 
By virtue of equations~\eqref{rec2}, however,
these time-Taylor coefficients can be written in terms of the lower-order 
time-Taylor coefficients $n-1$.
We thus can construct the time-Taylor coefficients of the displacement in 
a recursive way. We find in Fourier space ($n \geq 1$)
\begin{align}
 &\hat{\fett{\xi}}_{k_1}^{(n)} = - \ii \fett{k}\, \hat \phi^{\rm (init)} 
\delta_{1n} \nonumber \\ 
&\,\,- \left( k_1^2 + k_2^2 + k_3^2 \right)^{-2} 
    \Bigg(  \ii \fett{k} \frac{n-1/2}{2n+3} \left[ \hat \zeta^{(n-1)}_{k_1+1} + \hat \zeta^{(n-1)}_{k_1-1}   \right]  \nonumber \\ 
   &\,\, + \frac{n-2}{2n} \left(  -k_\perp^2 , \, k_1 k_2 ,\, k_1 k_3  \right)^{\rm T}
  \left[ \hat \chi^{(n-1)}_{k_1+1} + \hat \chi^{(n-1)}_{k_1-1}  \right] \!\!  \Bigg) \,,
\end{align}
and 
in real space
\begin{align} 
 &\fett{\xi}^{(n)} = - \nabla^{\rm L} \phi^{\rm (init)}\, \delta_{1n}  \nonumber \\
 &\,\,+\nabla_{\rm L}^{-2}  \Bigg( \!\! \frac{2n-1}{2n+3} \nabla^{\rm L}
  \left[  \zeta^{(n-1)} \cos q_1 \right] \nonumber \\
 &\,\,+\frac{n-2}{n} \left(  \upartial_2^{\rm L}\upartial_2^{\rm L}+
 \upartial_3^{\rm L}\upartial_3^{\rm L},\, -\upartial_1^{\rm L} \upartial_3^{\rm L} ,\, -\upartial_1^{\rm L} \upartial_{2}^{\rm L} \right)^{\rm T}
 \left[  \chi^{(n-1)} \cos q_1 \right]  \!\!  \Bigg) , \label{sol:displacement}
\end{align}
where $\nabla_{\rm L}^{-2}$ is the inverse Laplacian in Lagrangian coordinates.

From this we obtain, to order $\epsilon$ and to all orders in time, respectively the particle trajectory
and then, using \eqref{jacobepsilon}, the Jacobian, 
\begin{align}
  &x_i(\fett{q};\tau) = q_i - \delta_{1i} \tau \sin q_1 + \epsilon 
    \sum_{n=1}^\infty \xi_i^{(n)}(\fett{q})\, \tau^n \,, \label{pertMap} \\
  &J = 1 - \tau \cos q_1 + \epsilon \sum_{n=1}^\infty \! \left( \chi_{,1}^{(n)} 
    + \left[ 1 - \tau \cos q_1 \right] \zeta^{(n)} \right) \!\tau^n \,, \label{pertJac}
\end{align}
where $\chi_{,1}^{(n)} = \xi_{1,1}^{(n)}$ 
and $\zeta^{(n)} = \xi_{2,2}^{(n)}+\xi_{2,2}^{(n)}$.
Equations~\eqref{pertMap}--\eqref{pertJac} constitute  the main 
technical results of this paper.

In the following section we show that these formal solutions are actually
convergent series and free of any singularities. Then, in section~\ref{sec:tstar}, 
we comment on $\tau_\star$, the time of first shell-crossing in the perturbed problem.

\section{No Singularities in Lagrangian solutions}\label{sec:nosing}

Observe that in \eqref{evo1}, there is a term $[1- \tau \cos q_1] \R \zeta$, 
involving the second-order time derivative of one of the unknowns,  $\zeta$,
whose coefficient is the Jacobian of the unperturbed problem, i.e., $J^{(0)} =   1- \tau \cos q_1$. 
This term vanishes at the first shell-crossing for the
unperturbed (1D) problem. As it is known from e.g.\ Fuchsian theory \citep{Moser59}, 
the vanishing of the coefficient in front of the highest-derivative 
term may easily lead to a singularity (at least for ODEs). Should this happen
here, we would have to face a singular perturbation problem. Fortunately,
this is not the case, and the Lagrangian map determined by the first-order
perturbation equations \eqref{evolutioneqs}, is an  entire
function of time, as we now show.  Here, we must stress that \textit{after} shell-crossing, 
because of multi-streaming,  
the true Lagrangian map ceases to be governed by the Euler--Poisson 
equations \eqref{fluidequations}, 
but this does not matter for the determination
of the first shell-crossing.

It is easily shown that the entire character of the time-Taylor series is related
to the behaviour at large orders $n$ of the Taylor coefficients 
$\chi^{(n)}(\fett{q})$ and $\zeta^{(n)}(\fett{q})$. These satisfy the recursion
relations~\eqref{explicitrec}, which can be simplified for large $n$ and
approximated by their asymptotic form:
\begin{subequations} \label{rec1as}
\begin{align}
 &\chi_{,1}^{(n)} + \zeta^{(n)}   =   \zeta^{(n-1)} \cos q_1 \,, \label{rec1aschi}\\
 &\chi^{(n)} + k_\perp^{-2} \zeta_{,1}^{(n)} =   \chi^{(n-1)} \cos q_1 \,.\label{rec1aszeta}
\end{align}
\end{subequations}

The large-$n$ recursion relations~\eqref{rec1as} can actually be solved explicitly. 
Paradoxically, to achieve this, we shall return to $\tau$ space rather than working directly
with the time-Taylor coefficients and their recursion relations. 
However, we shall not work with the full Taylor series but only with lower-truncated series, i.e.,
\be
  \chi_{_N}(\fett{q};\tau) = \sum_{n=N}^\infty \chi^{(n)}(\fett{q}) \,\tau^n \,, \qquad 
  \zeta_{_N}(\fett{q};\tau) = \sum_{n=N}^\infty \zeta^{(n)}(\fett{q})\, \tau^n \,,
\label{timetaylorlt}
\ee
where $N$ is taken large enough to be able to use the asymptotic
form \eqref{rec1as} of the recursion relations.
Comparison of \eqref{timetaylor} and \eqref{timetaylorlt} shows that
$\chi_{_N}(\fett{q};\tau)$ and $\chi(\fett{q};\tau)$ differ by a polynomial in $\tau$
(with $\fett{q}$-dependent coefficients) of degree $N-1$. The same statement
holds for $\chi_{_N}(\fett{q};\tau)$ and $\chi(\fett{q};\tau)$. As a consequence,
it is equivalent to show the entire character in $\tau$ of the pair
$(\chi(\tau), \zeta(\tau))$ or of the pair $(\chi_{_N}(\tau), \zeta_{_N}(\tau))$.

At this point, to simplify the calculations, and without loss of generality,
we can set $k_\perp =1$ (if
not, rescale the transverse variables $q_\perp \equiv( q_2,\,q_3)$ and
the amplitude of the first component of the perturbation $\chi \equiv \xi_1$ suitably).

We now multiply~\eqref{rec1aschi} and~\eqref{rec1aszeta} by $\tau^n$ and
sum on $n$ from $N$ to infinity, to obtain the following equations:
\begin{subequations} \label{aschizeta}
\begin{align}
 &\chi_{_{N,1}} + (1-\tau \cos q_1)\zeta_{_N}   =   \tau^N \cos q_1\zeta^{(N-1)}\,, \label{aschi}\\
 &(1-\tau \cos q_1)\chi_{_N} + \zeta_{_{N,1}} =   \tau^N \cos q_1\chi^{(N-1)}\,.\label{aszeta}
\end{align}
\end{subequations}

The functions $\chi_{_N}$ and $\zeta_{_N}$ are coupled by a system of two
equations. However, by simply introducing the sum and the difference
\be
Z_N^\pm \equiv \chi_{_N} \pm \zeta_{_N}\,,
\label{sumdiff}
\ee
we obtain two decoupled equations:
\begin{equation}
Z^\pm_{N,1} = \mp[ 1 - \tau \cos q_1 ]Z^\pm_N + F^\pm_N(q_1;\tau) \,,
\label{ZNdiffeq}
\end{equation}
where 
\begin{equation}
F_N^\pm(q_1;\tau) \equiv  \tau^N \cos q_1 \left( \zeta^{(N-1)} \pm \chi^{(N-1)}\right)\,.
\label{deffpm}
\end{equation}
Observe that equations~\eqref{ZNdiffeq} are first-order ordinary
differential equations in $q_1$ in which the time appears just as a parameter; 
indeed, thanks to the large-$n$ asymptotics, $\tau$-derivatives have dropped out.
Also observe  that $F_N^\pm(q_1;\tau)$ are polynomials in $\tau$
and trigonometric polynomials in $q_1$, and thus entire functions of
$\tau$ and $q_1$.

We now show that equations~\eqref{ZNdiffeq} have unique solutions
within the class of functions of $q_1$ that are $2\uppi$-periodic. We begin with
the case of $Z_N^+ (q_1;\tau)$. 
For this, we first consider the (left spatial) initial value problem for
which we prescribe the value of $Z_N^+$ for some initial value $q_{\rm in}$ of
$q_1$ and look for the solution to its right, i.e. for $q\ge q_{\rm in}$. This has 
the explicit solution 
\begin{align}
&Z_N^+(q_1;\tau) = G^+(q_1,q_{\rm in}; \tau)\,Z_N^+(q_{\rm in})  \nonumber \\
&\qquad\hspace{1cm} +\int_{q_{\rm in}}^{q_1} \dd q' G^+(q_1,q'; \tau) \, F_N^+(q';\tau)\,, \label{explicitsol}
\end{align}
in terms of the Green's function $G^+$ of the associated linear differential
equation (without the $F_N^+$ term), given, for $q_1\ge q'$, by
\begin{align}
G^+(q_1,q';\tau) &= \exp \left\{ - \int_{q'}^{q_1} \dd q'' \left[ 1 -\tau \cos q'' \right]\right\} \nonumber  \\
 &=  \exp \left\{\tau \left(\sin q_1 -\sin q' \right)- \left( q_1-q'\right) \right\}\,.
\label{green}
\end{align}
The solution  \eqref{explicitsol} is in general not periodic in
$q_1$, but a periodic solution can be constructed by letting 
$q_{\rm in} \to -\infty$, because far to the right of the
point $q_{\rm in}$, the solution relaxes to spatial
periodicity. This is proved by decomposing the interval 
$[q_{\rm in},q_1]$ into adjacent intervals all-but-the-first 
(which may be smaller) of length $2\uppi$.  
The number $M$ of intervals of length $2\uppi$ is the
integer part of  $|q_1-q_{\rm  in}|/(2\uppi)$ and thus tends to
infinity when $q_{\rm in}\to -\infty$.  We observe that, 
when the argument $q'$ of the integrand on the right-hand side of~\eqref{explicitsol} 
is shifted from one interval to the neighbouring left interval by 
subtracting $2\uppi$, the integrand is multiplied by a
factor $\exp(-2\uppi)$. Indeed, the only $q'$-dependent term which
is not periodic in $q'$ is $(q_1-q')$ in the exponential, 
which generates the  stated factor.  Hence, the sum over all the 
$M$ intervals produces a geometric series of ratio  $\exp(-2\uppi)$. 
As $q_{\rm in} \to -\infty$ at fixed $Z_N^+(q_{\rm in})$, 
the first term on the right-hand side of~\eqref{explicitsol} tends to zero, 
and the second term is given by the sum
of an infinite convergent geometric series, namely
\begin{align}
Z_N^+(q_1;\tau) &=\frac{1}{1-\e ^{-2\uppi}}\int_{q_1-2\uppi}^{q_1} \dd q' F_N^+(q';\tau) \nonumber \\
   &\quad\hspace{0.5cm}\times \exp \{\tau (\sin q_1 -\sin q')- (q_1-q')\}\,.
\label{periodicsol}
\end{align} 
It is easily checked that \eqref{periodicsol} is a $2\uppi$-periodic
solution of~\eqref{ZNdiffeq}, and, furthermore, is the unique one. 
Indeed, let $Z_N^+(q_1;\tau)$ and $\tilde Z_N^+(q_1;\tau)$ be two such
solutions. Their difference $\Delta (q_1;\tau)$ satisfies the
homogeneous equation
\begin{equation}
\Delta_{,1} = (\tau \cos q_1 -1) \Delta\,,
\label{deltaeq} 
\end{equation}
which implies that 
\begin{equation}
\Delta(q_1+2\uppi\hspace{0.02cm}; \tau) = \e ^{-2\uppi} \Delta(q_1; \tau)\,.
\label{almosthere}
\end{equation} 
Since $Z_N^+(q_1;\tau)$ and $\tilde Z_N^+(q_1;\tau)$ are
$2\uppi$-periodic, so is their difference $\Delta$, which by~\eqref{almosthere}  
vanishes, hence we confirm the uniqueness of the solution.
Furthermore, using the fact that $F_N^+(q';\tau)$ is polynomial in $\tau$, 
we easily check that, for any real $q_1$, $Z_N^+(q_1;\tau)$, given by \eqref{periodicsol},
is an entire function of $\tau$.

The case of $Z_N^-(q_1;\tau)$, which satisfies \eqref{sumdiff} with the minus sign,
is handled similarly, except that we must replace the left spatial initial
value problem by a right spatial initial value problem, where we seek the solution for $q\le q_{\rm in}$ 
(or, equivalently, we can just change $q_1$ into $-q_1$). 
Hence  $Z_N^-(q_1;\tau)$ is also an entire function of $\tau$.
As a consequence $\chi(q_1; \tau)$  and $\zeta(q_1; \tau)$ are, for any real $q_1$, 
entire functions of $q_1$. 
It is also easily shown that they are also entire functions of $\tau$.

\section{The time of perturbed shell-crossing} \label{sec:tstar}

In the absence of perturbations, when the flow is exactly one-dimensional and with our initial condition, 
shell-crossing happens at the time $\tau_\star^{(0)}=1$ and location $q_{1\star} =0$,
and for arbitrary $q_2$ and $q_3$. Thus, the whole plane $q_1=0$ shell-crosses at $\tau_\star^{(0)}=1$. 
When the perturbation is switched on, translation invariance in the directions of $q_2$ and $q_3$ 
is broken and, generically shell-crossing takes place at 
a time $\tau_\star \neq 1$ and at a single location $(q_{1\star},\, q_{2\star},\,q_{3\star})$. 
We shall now show that $\tau_\star$ is generically happening earlier than $\tau_\star^{(0)}=1$ 
and explain how the precise time and location can be obtained. 

For this purpose we will assume that the initial perturbation is a finite-order
trigonometric polynomial  
\begin{align} \label{trigo}
 \epsilon\, \phi^{\rm (init)} &= \epsilon \sum_{m=1}^M\sum_{n=1}^N \left[ a_{mn}^{\rm (init)}(q_1) \cos (mq_2 +nq_3)\right. \nonumber \\
 &\qquad \hspace{1.5cm} + \left.b_{mn}^{\rm (init)}(q_1) \sin(m q_2+n q_3) \right] \,.
\end{align}
Infinite Fourier series can also be handled, but this requires some
functional analysis which we would rather avoid here.

Because of the linearity of  \eqref{scalarEpsilon}--\eqref{eq:epsCauchy3} and of their autonomous character 
in $q_2$ and $q_3$, it is enough to know how to solve the linearized equations with an initial condition given 
by a single term in the sum \eqref{trigo}. As we shall see, the solution is needed only at
$\tau =1$; it can be obtained by summing the time-Taylor series \eqref{timetaylor} to a suitable order, 
depending on the desired accuracy (6--8th order is usually more than enough). In this way, one 
obtains  the following expression for the  first-order perturbation of the displacement:   
\begin{align}
 \epsilon\, \fett{\xi} 
  &= \epsilon\sum_{m=1}^M \sum_{n=1}^N \Big[ \fett{\mathfrak{a}}_{mn}(q_1;\tau) \cos (mq_2 +nq_3)\Big. \nonumber \\
  &\qquad \hspace{1.95cm}+ \Big.\fett{\mathfrak{b}}_{mn}(q_1;\tau) \sin (mq_2+nq_3) \Big]\,.
  \label{xifirstorder}
\end{align}
From this, one can calculate the Jacobian up to first order in $\epsilon$ to
obtain
\be
J= 1-\tau \cos q_1 +\epsilon[ (1-\tau \cos q_1)(\xi_{2,2}+\xi_{3,3}) +\xi_{1,1}]\,.
\label{Jfirstorder}
\ee
As we know, for $\epsilon =0$ the Jacobian vanishes for the first time at 
$\tau_\star^{(0)} =1$ and $q_{1\star}^{(0)} =0$ (the coordinate system was actually chosen to ensure this,
without loss of generality for 1D flow). For small~$\epsilon$, by continuity, the
perturbed  Jacobian will vanish at a time and place close to $\tau=1$ and $q_1=0$. For such values, we have
\be
 1-\tau \cos q_1 \approx (1-\tau) +\frac{q_1^2}{2} + {\rm h.o.t.}\,, \label{J0expand}
\ee
where h.o.t.\ stands for higher-order terms in $1-\tau$ and $q_1$. To determine
the leading order of the perturbed first shell-crossing we may thus discard
in \eqref{Jfirstorder} the higher-order term involving the factors $\epsilon (1-\tau \cos q_1)$ 
and use the following approximation:
\be
J \approx (1-\tau) +\frac{q_1^2}{2} + \epsilon \,\xi_{1,1}(0,q_2,q_3;1) + {\rm h.o.t.}\,,
\label{leadingJ}
\ee
where $\xi_{1,1}(0,q_2,q_3;1)$ denotes the $q_1$ derivative of the first
component of the displacement $\xi$, evaluated at time $\tau =1$, at $q_1=0$
and arbitrary $q_2$ and $q_3$ (so far). Using \eqref{xifirstorder}, the Jacobian takes the following form
\begin{align}
J &\approx (1-\tau) +\frac{q_1^2}{2} + \epsilon \sum_{m=1}^M\sum_{n=1}^N
\left[\alpha_{mn} \cos (mq_2+nq_3)  \right. \nonumber \\
&\qquad \hspace{2cm} +\left. \beta_{mn} \sin (mq_2+nq_3)\right] +
{\rm h.o.t.}\,,
\label{leadingexplicitJ}
\end{align}
where the coefficients $\alpha_{mn}$ and $\beta_{mn}$ are easily expressed
in terms of the coefficients $\mathfrak{a}_{mn}$ and $\mathfrak{b}_{mn}$ that
appear in~\eqref{xifirstorder}.

Although the determination of $\mathfrak{a}_{mn}$ and $\mathfrak{b}_{mn}$ cannot
be done by purely analytic means, the very form of \eqref{xifirstorder} allows
us to conclude that the first shell-crossing takes place at time
$\tau_\star = 1+\epsilon C$, where
\begin{align}
C &\equiv  \min_{q_2,q_3} \sum_{m=1}^M\sum_{n=1}^N
\left[\alpha_{mn} \cos (mq_2+nq_3)  \right. \nonumber \\
 &\qquad \hspace{2.6cm} +\left. \beta_{mn} \sin (mq_2+nq_3)\right]\,.
\label{cmin}
\end{align}
The first shell-crossing is at $q_{1\star} =0$ and at those
values of $q_{2\star}$ and $q_{3\star}$ for which the minimum in~\eqref{cmin}
is achieved.
That this \mbox{minimum} is negative, and thus that \textit{the
perturbed shell-crossing takes place slightly before the unperturbed
one}, follows from the observation that a trigonometric polynomial (in
one or several variables) without a constant term necessarily takes
both positive and negative values, because it is continuous and its
space average over the period(s) is zero.

\section{A concrete example: the three sine waves model}\label{3waves}

Let us apply the developed tools to a concrete example, for which we 
determine, to first order in the perturbation, the Jacobian of the Lagrangian map 
and the time of first shell-crossing. 
We set the initial gravitational potential to 
\be
\varphi^{\rm (init)}  = -\cos q_1 +   \epsilon_2 \sin q_2 + \epsilon_3 \sin q_3 \,.
\label{initial3}
\ee
Here, $\epsilon_2 = \epsilon\, C_2$ and $\epsilon_3 =\epsilon\, C_3$  
with $\epsilon>0,\, C_2>0,\, C_3>0$, so that the relative amplitudes of the $q_2$-dependent
perturbation and  of the $q_3$-dependent
perturbation can be taken arbitrary.

From the analysis of the previous section, obtaining the time of first shell-crosssing requires
the knowledge of the perturbed Jacobian at $\tau =1$ and $q_1 =0$. The latter is given by~\eqref{pertJac}
which involves a  time-Taylor series to all orders in $\tau$, which is guaranteed to converge at $\tau =1$
because we are dealing, as we have seen in section~\ref{sec:nosing}, with an entire function of $\tau$.
In practice, to obtain numerical approximations for the time of first shell-crossing, we can
truncate this time-Taylor series to a finite order ${\cal N}$, using instead of~\eqref{pertJac}, the
truncated Jacobian
\be
  J_{\cal N}= 1-\tau\cos q_1 + \epsilon\sum_{n=1}^{\cal N} \! \left( \chi_{,1}^{(n)} 
    + \left[ 1 - \tau \cos q_1 \right] \zeta^{(n)} \right) \!\tau^n  \,.\label{quantity}
\ee
Given that the initial perturbation~\eqref{initial3} is composed of two sine waves in the transverse
coordinates $q_2$ and $q_3$, the Jacobian and its truncations will also have this property and
we can write
\begin{align}
 J_{{\cal N}} &= 1 - \tau \cos q_1  \nonumber  \\
  &\qquad+ \sum_{n=1}^{\cal N} \kappa^{(n)} (q_1;\tau) 
   \left[ \epsilon_2 \sin q_2 + \epsilon_3 \sin q_3  \right] \tau^n  \,, \label{Jsinewaves}
\end{align}
where the  coefficients $\kappa^{(n)} (q_1;\tau)$ are easily computed by using our recursion 
relations~\eqref{sol:displacement} and symbolic algebra tools.
The first few coefficients to order ${\cal N}$ read 
\begin{align}
   &\kappa^{(1)}=  1 - \tau \cos q_1   \,, \\
  &\kappa^{(2)} =  \frac{3}{14} (2 -\tau  \cos q_1) \cos q_1   \,, \\
  &\kappa^{(3)} = \frac{\cos q_1}{420}  \left[ 50 \cos q_1 + \tau  \left\{ \cos (2 q_1)-25\right\} \right]  \,, \\
  &\kappa^{(4)} = \frac{\cos q_1}{184800} \Big[7000 -280 \cos (2 q_1) \nonumber \\
  &\qquad \hspace{1.8cm}-2715 \tau  \cos q_1+443 \tau  \cos (3 q_1) \Big] \,. 
\end{align}
Higher orders are somewhat bulky and are given in Appendix~\ref{app:sine}  (for $5\leq {\cal N} \leq 8$).

To obtain the perturbed time of first shell-crossing to leading order in $\epsilon$ 
we can set $\tau=1$  and $q_1=0$ in the coefficients $\kappa^{(n)}$, because the discrepancies will only contribute
to higher orders. In this way, we finally obtain the following Taylor-truncated approximations
\be
\tau_\star \simeq 1 -  c_{\cal N} \left( \epsilon_2+\epsilon_3 \right) \,,
\label{taustarN}
\ee
where
\be
c_4 = 0.3003,\quad c_6 = 0.3073,\quad c_7 = 0.3076,\quad c_8 = 0.3077\,.
\label{cNapprox}
\ee
As to the spatial location of this first shell-crossing, 
it is found to be at $q_{1\star} = 0$
and $q_{2\star} = q_{3\star} = -\uppi/2$ (modulo $2\uppi$).

\section{Concluding Remarks} \label{sec:conclusions}

 For any time-Taylor series, the radius of convergence is the distance
between the expansion point and
the closest singularity in the complex-time plane.
Applying this statement to the Euler--Poisson equations 
{\it in a Eulerian formulation}, 
where the density (and velocity) is expanded in a time-Taylor series,
it is evident that the radius of convergence cannot be infinite,
neither in 1D nor beyond, 
because of the explicit appearance of real-space density singularities at shell-crossing. 
In a Lagrangian formulation, by contrast, 
the use of the Lagrangian map acts as the desingularization transformation of the problem, 
and thus, shell-crossing can be investigated in a rigorous way.
In particular, we have shown that in Q1D, the time-Taylor coefficients of
the \textit{linearized displacement field} are all
non-vanishing, but we also found that its Taylor series has an infinite radius of convergence. 
Without linearization this is unlikely to remain true, and
we expect that there will be complex-time singularities and a finite
radius of convergence of the Taylor series, which will however be large when
the perturbation parameter $\epsilon$ is small.

Time and location of the first shell-crossing -- which in Lagrangian coordinates is not a singularity but just a vanishing of the Jacobian,  
can be found by perturbation theory. In practice,
one needs to calculate a sufficient number of
time-Taylor coefficients $\fett{\xi}^{(n)}$ for the 
displacement field 
$\fett{\xi}=\sum_{n=1}^\infty \fett{\xi}^{(n)} \tau^n$, which are easily 
generated by the use of our novel recursion relations~\eqref{sol:displacement}.
As a general rule of thumb --
verified for various initial conditions of 
the type~\eqref{trigo} -- 
the time of shell-crossing can be determined to four digit accuracy 
when truncating the time-Taylor expansion of the displacement up to order $n = 7$.

In our Lagrangian formulation, the classical ZA is achieved by
setting \mbox{$n=1$} in the time-Taylor series, 
and discarding all higher-order time-Taylor coefficients.
Tradionally, the ZA has not only been applied to the purely 1D
but also to the 2D and 3D case. 
Most famous in this context is the prediction of the so-called
Zel'dovich pancake that originates from the gravitational collapse of 
an ellipsoidal distribution of pressureless matter (see, e.g., \citealt{ASZ:1982}).
In one dimension the ZA is exact until shell-crossing, so one could hope that in a quasi-one-dimensional situation, the ZA would still give meaningful results.
This, however, need not be the case, as is now demonstrated  with a simple 
counterexample. Setting $\varphi_{\rm g}^{(\rm init)} =  -\cos q_1 + \epsilon \sin q_2$, 
it is straightforward to determine the particle trajectory within the classical ZA, 
it reads $x_i^{\rm ZA} = q_i -  \tau \delta_{i1}\sin q_1  - \epsilon  \tau \delta_{i2}\cos q_2$.
Analysing the entries of the Jacobian matrix $x_{i,j}^{\rm ZA}$, it is seen that, within the ZA and to any order in $\epsilon$, 
shell-crossing happens at $\tau_\star^{\rm ZA} =1$ and at $q_{1\star}^{\rm ZA} = 0$ 
(modulo $2\uppi$) and arbitrary values  of $q_2$ and $q_3$.
However, using our tools and determining the displacement up to the 8th order in
the time-Taylor series, we find that shell-crossing happens already 
at $\tau_\star = 1 - 0.3077 \epsilon$, at the position~${q}_{1\star}= 0$ and $q_{2\star} = -\uppi/2$ (and arbitrary $q_3$). 
Thus, shell-crossing happens earlier than predicted by the ZA, 
and furthermore it occurs at a specific value of $q_{2\star}$, and not for 
arbitrary values of $q_2$.

We note that in our model we have assumed that quasi-one-dimensionality holds already  at 
initial time $\tau = 0$. An improved model would have quasi-one-dimensionality holding after a pancake has formed at
some $\tau^{\rm (init)} >0$.
To handle this one should use time-Taylor expansions around  $\tau^{\rm (init)}$ (and not around $\tau =0$). 
Since the perturbed Euler--Poisson equations~\eqref{eqs:epsilon} are non-autonomous in the time variable, 
it follows that the resulting time-Taylor coefficients around the ``shifted'' expansion point will differ 
from the ones we have obtained. Developing recursion relations for the time-shifted expansion is in principle
fairly straightforward, but will be left for future work.

What happens after shell-crossing? Although this is an important
question, in the present paper we have focused on the time before and
at shell-crossing. We thus leave this issue to follow-up studies.
Qualitatively, it is expected that for sufficient short times after
shell-crossing, the fluid description should still deliver physically
meaningful results, provided that the Poisson equation and Lagrangian
mass conservation are appropriately generalized -- to take into account
the multiple branches of the Lagrangian map. Deep into the multi-stream
regime, however, a phase-space description becomes eventually
mandatory. Such a phase-space description, in Q1D and beyond, is
still missing in the literature (for the 1D case, see
\citealt{Colombi:2014zga,Colombi:2014lda,Taruya:2017ohk}, and for
approximative models beyond 1D, see \citealt{Buchert:1997dr}).
In principle,
cosmological $N$-body simulations aim to solve the said phase-space
dynamics to high accuracy, however, being a brute-force method by
nature and furthermore relying on a particle description, it is quite
a challenge to gain mathematical insight about shell-crossing and the time
after. In this context, we note the novel cosmological simulations of
\citet{Hahn:2012ma,Hahn:2015sia}, where $N$-body particles are used as
tracers of (adaptively refineable) phase-space elements. Here, a smooth
representation of the gravitational field is obtained, which improves
the force computation especially near caustics, and delivers the
phase-space dynamics to a good approximation.

Finally, let us comment on the possibility of applying our methodology
to the relativistic shell-crossing in Q1D, an outstanding problem
within the field of General Relativity. For reasons similar to those
given in this paper, we expect that a Lagrangian-coordinates approach
would be more fruitful than an Eulerian one. In particular, the use of the
Lagrangian map could possibly desingularize the relativistic problem
as well. For an irrotational and pressureless matter fluid, it is
known that a synchronous-comoving coordinate system resembles the
relativistic Lagrangian frame of reference \citep[see
e.g.][]{Rampf:2014mga,Rigopoulos:2014rqa}. A promising starting point for such an
investigation could be the relativistic Lagrangian
equations~(24)--(27) of \citet{Alles:2015vua}, which are closely
related to our starting point, the Newtonian Lagrangian
equations~\eqref{eqs:lag}.

\section*{Acknowledgements}

We thank Tom Abel, Thomas Buchert, St\'ephane Colombi, Vincent Desjacques, Adi Nusser, Fabian Schmidt, 
Sergei Shandarin, and Barbara Villone for useful discussions and/or comments on the manuscript.
CR is supported by the DFG through the SFB-Transregio TRR33 ``The Dark Universe''. 
CR thanks the Lagrange Laboratory of the Observatoire de la C\^ote d'Azur
and UF thanks the Technion Department of Physics for their hospitality and support.



\bibliographystyle{mnras}
\bibliography{desing} 



\appendix

\section{Higher-order Taylor coefficients for the three sine waves model}\label{app:sine}

In section~\ref{3waves} we have applied our tools to an explicit example, 
the so-called three-sine waves model, for which we have generated solutions for the Jacobian up to
order ${\cal N}=8$ (see equation~\eqref{Jsinewaves}). 
For brevity, we have skipped in the main text the terms beyond order ${\cal N} >4$; 
for the higher-order terms we find
\begin{align}
  &\kappa^{(5)} = \frac{\cos q_1 }{680680000}  \Big[17 \big\{ 76316 \tau  \cos (2 q_1) +407250 \cos q_1 \nonumber \\
   & -75 (886 \cos (3 q_1)+2715 \tau ) \big\}+59485 \tau  \cos (4
   q_1) \Big]  \,, \\
  &\kappa^{(6)} = -\frac{\cos q_1}{2654652000000}  \Big[ \tau  \big\{ 2659396870 \cos
   (q_1) \nonumber \\
 &-962623441 \cos (3 q_1)  +28312075 \cos (5 q_1) \big\} \nonumber \\
 &+2860 \big( 1297372
   \cos (2 q_1)+59485 \cos (4 q_1)-3461625 \big) \Big]  \,,  
\end{align}
\begin{align}
&\kappa^{(7)} = \frac{10^{-7} \cos q_1}{198703356852}  \Big[ 37 \big\{ 3 \tau  \big( 3999222278389 \cos (2
   q_1) \nonumber \\
  &-55818626690 \cos (4 q_1) \big) +15470 (-962623441 \cos (3
   q_1) \nonumber \\
  &+28312075 \cos (5 q_1)  -1329698435 \tau ) \nonumber \\
 &+41140869578900 \cos q_1 \big\}
  -1201960568875 \tau  \cos (6 q_1) \Big] \,, \\
&\kappa^{(8)} = \frac{10^{-8} \cos q_1}{130879277713184} \Big[ \tau  \big\{ 407511682759452297 \cos (3 q_1)  \nonumber \\
  &-827808374798014805 \cos q_1 +335214948722225 \cos (7 q_1) \nonumber \\
  & -32682412580801125 \cos (5 q_1) \big\} +5200 \big( 761106087209650 \nonumber \\
  &-443913672901179 \cos (2q_1)
   +6195867562590 \cos (4 q_1) \nonumber \\
  &+1201960568875 \cos (6 q_1) \big) \Big] \,.
\end{align}

\bsp	
\label{lastpage}
\end{document}